\documentclass[aps,prx,reprint,superscriptaddress]{revtex4-2}

\usepackage{graphicx}
\usepackage{mhchem}

\usepackage{amsmath,amssymb,amsfonts,bm,graphicx,mathtools,braket}
\usepackage{hyperref}
\usepackage{xcolor}	

\newcommand{\unit}[1]{\,\mathrm{#1}} 
\newcommand{\equa}[1]{Eq.~\eqref{#1}} 
\newcommand{\fig}[1]{Fig.~\ref{#1}}

\usepackage{ulem}

\begin{document}
	
	\title{Snapshots of a light-induced metastable hidden phase driven by the collapse of charge order}
	
	
	\newcommand{\MITCHEM}{Department of Chemistry, Massachusetts Institute of Technology, Cambridge, Massachusetts, USA, 02139}
	\newcommand{\MITPHYS}{Department of Physics, Massachusetts Institute of Technology, Cambridge, Massachusetts, USA, 02139}
	\newcommand{\MITECE}{Department of Electrical Engineering and Computer Science, Massachusetts Institute of Technology, Cambridge, Massachusetts, USA, 02139}
	\newcommand{\AUSPHYS}{Department of Physics, The University of Texas at Austin, Austin, Texas, USA, 78712}
	\newcommand{\HARVPHYS}{Department of Physics, Harvard University, Cambridge, Massachusetts, USA, 02138}
	
	\author{Frank Y. Gao}
	\altaffiliation{These authors contributed equally to this work}
	\affiliation{\MITCHEM}
	\author{Zhuquan Zhang}
	\altaffiliation{These authors contributed equally to this work}
	\affiliation{\MITCHEM}
	\author{Zhiyuan Sun}
	\affiliation{\HARVPHYS}
	\author{Linda Ye}
	\altaffiliation{Present address: Department of Applied Physics, Stanford University, Stanford, California, USA, 94305 }
	\affiliation{\MITPHYS}
	\author{Yu-Hsiang Cheng}
	\altaffiliation{Present address: Department of Electrical Engineering, National Taiwan University, Taipei, Taiwan, 10617}
	\affiliation{\MITECE}
	\author{Zi-Jie Liu}
	\affiliation{\MITCHEM}
	\author{Joseph G. Checkelsky}
	\affiliation{\MITPHYS}
	\author{Edoardo Baldini}
	\email{edoardo.baldini@austin.utexas.edu}
	\affiliation{\AUSPHYS}
	\author{Keith A. Nelson}
	\email{kanelson@mit.edu}
	\affiliation{\MITCHEM}

	
	\date{\today}
	
	\begin{abstract} 
		Nonequilibrium hidden states, both transient and long-lived, provide a unique window into thermally inaccessible regimes of strong coupling between microscopic degrees of freedom in quantum materials.  Understanding the physical origin of these states is of both fundamental and practical significance, allowing the exploration of far-from-equilibrium thermodynamics and the development of optoelectronic devices with on-demand photoresponses. However,  mapping the ultrafast formation of a  long-lived hidden phase  remains a long-standing challenge in physics since the initial state of the system is not recovered rapidly and conventional pump-probe methods are thus not applicable.  Here, using a suite of state-of-the-art single-shot spectroscopy techniques, we present a direct ultrafast visualization of the photoinduced phase transition to both transient and long-lived hidden states in an electronic crystal, 1\textit{T}-TaS$_2$.  Capturing the dynamics of this  complex phase transformation in a single-shot fashion demonstrates a commonality in microscopic pathways, driven by the collapse of charge order, that the system undergoes to enter the hidden state and provides unambiguous spectral fingerprints that distinguish  such state from thermally accessible phases. We present a theory of fluctuation dominated process that explains both the dynamics and the nature of the metastable state. Our results settle the debate around the origin of this elusive metastable state and pave the way for the discovery of new quantum phases of matter. 
	\end{abstract}
	
	
	\maketitle
	
	\section{Introduction}
	Ultrafast light-matter interactions can trigger a plethora of  exotic phenomena in quantum materials \cite{zhang2014dynamics, basov2017towards}, such as light-induced superconductivity \cite{cremin_photoenhanced_2019, fausti_light-induced_2011, mitrano_possible_2016}, nonlinear phononic control of lattices \cite{forst_nonlinear_2011, nova_effective_2017, li_terahertz_2019} and photon-dressed topological phases \cite{wang_observation_2013, mahmood_selective_2016}. In addition, the study of photoinduced hidden phases \cite{stojchevska_ultrafast_2014, zhang_cooperative_2016}, i.e., states inaccessible in equilibrium phase diagrams, has recently emerged as a new field of research. While many hidden phases that are induced by laser pulses are short-lived \cite{morrison2014photoinduced, otto2019optical, li_terahertz_2019, shi2019ultrafast}, a few such phases can persist indefinitely under suitable environmental conditions \cite{stojchevska_ultrafast_2014, zhang_cooperative_2016,mcleod2020multi,liu_photoinduced_2021}. For such nonequilibrium  metastable phases, salient gaps in our understanding remain.  Conventional pump-probe spectroscopy methods, being stroboscopic in nature, can provide profound insights into many phase transitions but are not applicable when the material does not return to its initial state after each pump laser shot. Single-shot time-resolved spectroscopy techniques, on the other hand, can capture, with a single pump laser shot, the real-time dynamical evolution of slowly reversible and irreversible processes.\cite{zhang_cooperative_2016, teitelbaum_real-time_2018, teitelbaum_dynamics_2019} {Single-shot} measurements {thereby can} offer unique mechanistic insights into the genesis of metastable hidden phases.
	
	A metastable hidden phase {whose origin and formation pathway are highly debated \cite{stojchevska_ultrafast_2014,ritschel2015orbital,stahl2020collapse}} occurs in the {prototype quasi-two-dimensional charge-density wave (CDW) crystal, 1\textit{T}-TaS$_2$. The crystal} undergoes successive first-order phase transitions upon cooling. First, from the high-temperature incommensurate CDW state (IC state), 1\textit{T}-\ce{TaS2} enters a nearly commensurate CDW state (NC state) at 350 K. It then undergoes a phase transition to a commensurate state (C state) at 180 K, below which a Mott gap develops.\cite{wilson1975charge,sipos2008mott} In the C state, an ultrafast photoinduced phase transition to a metastable hidden state (H state) {induced} by a single laser pulse has been observed at low temperature ($<10$ K), accompanied by a drop in the resistance of several orders of magnitude.\cite{stojchevska_ultrafast_2014} {The} highly conductive state can be erased by increasing the  sample temperature or annealing  thermally with a train of stretched pulses but is otherwise persistent. Due to its intriguing properties and the potential memory applications {of this type of photoinduced phase,} {the} metastable H state {of 1\textit{T}-TaS$_2$} has been investigated by transport {measurements}\cite{vaskivskyi_controlling_2015, vaskivskyi_fast_2016}, stroboscopic pump-probe spectroscopy \cite{stojchevska_ultrafast_2014, ravnik2021time}, scanning tunneling microscopy \cite{gerasimenko2019intertwined, ravnik2021time}, transmission electron microscopy \cite{sun_hidden_2018} and X-ray diffraction \cite{stahl2020collapse}. However, all the measurements {of this persistent phase} examine observables characterized before and after switching and thereby cannot provide insights into the initial dynamics of H state formation. In contrast, at higher temperatures (e.g., $>70$ K), it is argued that the metastable H state can also be induced, but only transiently as the system reverts to the pristine C state between shots. {Using a three-pulse pump-probe technique, at such sample temperatures, a first pulse of sufficient fluence was used to induce a phonon frequency shift characteristic of the H phase that was observed by variably delayed pump-probe measurements.\cite{ravnik_real-time_2018,ravnik2021time} Since the metastable H state formed at low temperature is persistent and the one formed at higher temperatures is transient, it remains an open question as to whether these states are essentially the same. Does the photoinduced switching in these two different temperature regimes follow distinct or common microscopic pathways? How do macroscopic properties like optical conductivity in these states differ from those in equilibrium?} Addressing these questions has been a longstanding challenge due to the lack of appropriate tools for tracking the {ultrafast} formation of long-lived hidden phases. Here, by conducting dual-echelon single-shot time-resolved spectroscopy experiments in both the near-infrared (NIR) and terahertz (THz) spectral ranges (Fig. 1)\cite{shin2014dual}, we directly capture the light-induced nucleation, stabilization, conductivity, and relaxation {associated with} {the transformation into the transient or persistent phase} and unveil the fundamental connections between the {photoinduced states and the pathways into them}.
	
	\begin{figure}
		\includegraphics[width=\columnwidth]{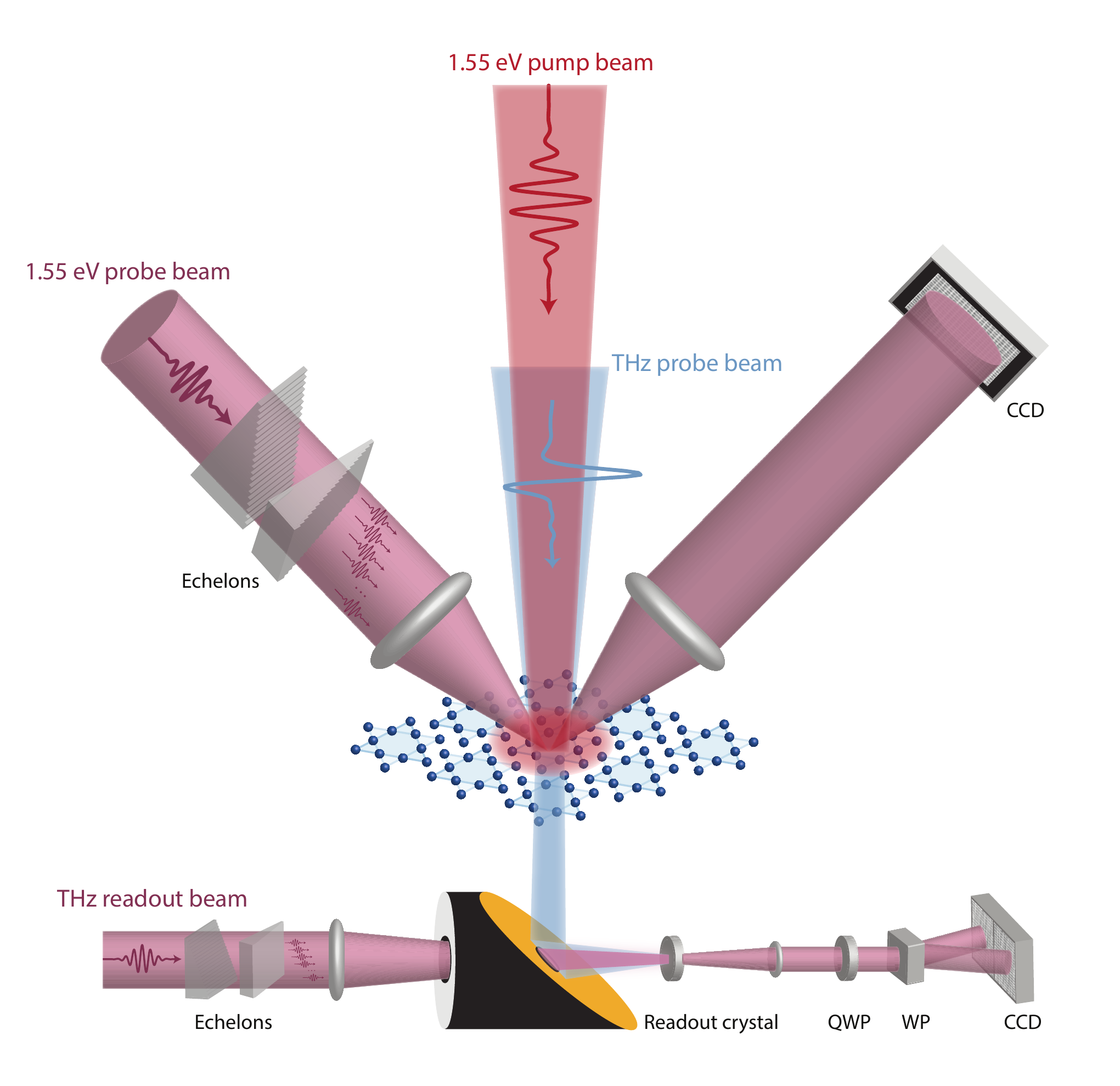}%
		\caption{\label{fig:Fig1}Schematic illustration of dual-echelon single-shot optical reflectivity and THz transmission spectroscopy. The sample is photoexcited by a 1.55 eV pump laser pulse. In single-shot reflectivity, the {1.55 eV} probe beam {(incident from top left)} is passed through a set of dual 20-step echelons and split {into a 20 x 20 grid of 400 pulselets with different time delays}. These probe pulselets are focused onto the 1\textit{T}-\ce{TaS2} sample along with the 1.55 eV pump pulse (red). The reflected probe pulselets are detected on different regions of a CCD {camera}. In single-shot THz transmission measurements, {a THz beam is incident on the sample (from above) and the transmitted THz field is focused into an electro-optic {readout} crystal in which field-induced birefringence gives rise to time-dependent optical polarization rotation proportional to the instantaneous THz field amplitude. A 1.55 eV readout beam (incident from bottom left) is passed through the echelons to generate 400 pulselets that overlap with the THz field in the EO crystal at different times. The transmitted pulselets are} passed through a quarter-wave plate (QWP)  and {a} Wollaston prism (WP), {producing two 20 x 20 grids of beams for balanced detection at a CCD camera.} In both cases, the resulting grid images are binned and unfolded to obtain a 9.3 ps time-domain trace of transient reflectivity or THz signal all in a single laser shot. }
	\end{figure}
	
	\section{Results}
	
	\subsection{THz conductivity of the H state}
	We first establish the creation of the H phase in 1\textit{T}-TaS$_2$ upon photoexcitation with a single, intense laser pulse at 1.55 eV (fluence $F= 2.5$ mJ/cm$^2$). To this aim, we measure the THz optical conductivity in the steady state before and after excitation at 7.8 K. Before irradiation (blue curve in Fig. 2(a)), the value of the real part $\sigma_1$ ($\sim$5 $\Omega^{-1}$ cm$^{-1}$) and its featureless spectrum are consistent with  the presence of a fully gapped insulating state.\cite{dean_polaronic_2011} Pumping the material results in a drastic change of the steady-state conductivity (red curve in Fig. 2(a)): the conductivity reaches values of $10^3 \hspace{3pt} \Omega^{-1} \text{cm}^{-1}$, which is about an order of magnitude larger than its counterpart in the NC CDW phase at 230 K.\cite{dean_polaronic_2011} This more conducting behavior is one of the fingerprints of the metastable H phase, and is consistent with previous transport measurements in the DC limit.\cite{vaskivskyi_controlling_2015, vaskivskyi_fast_2016} Unlike DC transport, THz spectroscopy has access to the conductivity over a wide spectral range, providing insights into the nature of the charge carriers. We observe that the value of $\sigma_1$ is relatively energy-independent above 2 meV and slightly suppressed below. This response is not captured by the conventional Drude model, suggesting a suppression of long-range transport due to electronic or structural disorder.\cite{ulbricht_carrier_2011} 
	
	\subsection{Single-shot transient reflectivity of the persistent H state}
	
	\begin{figure}
		\includegraphics[width=0.9\columnwidth]{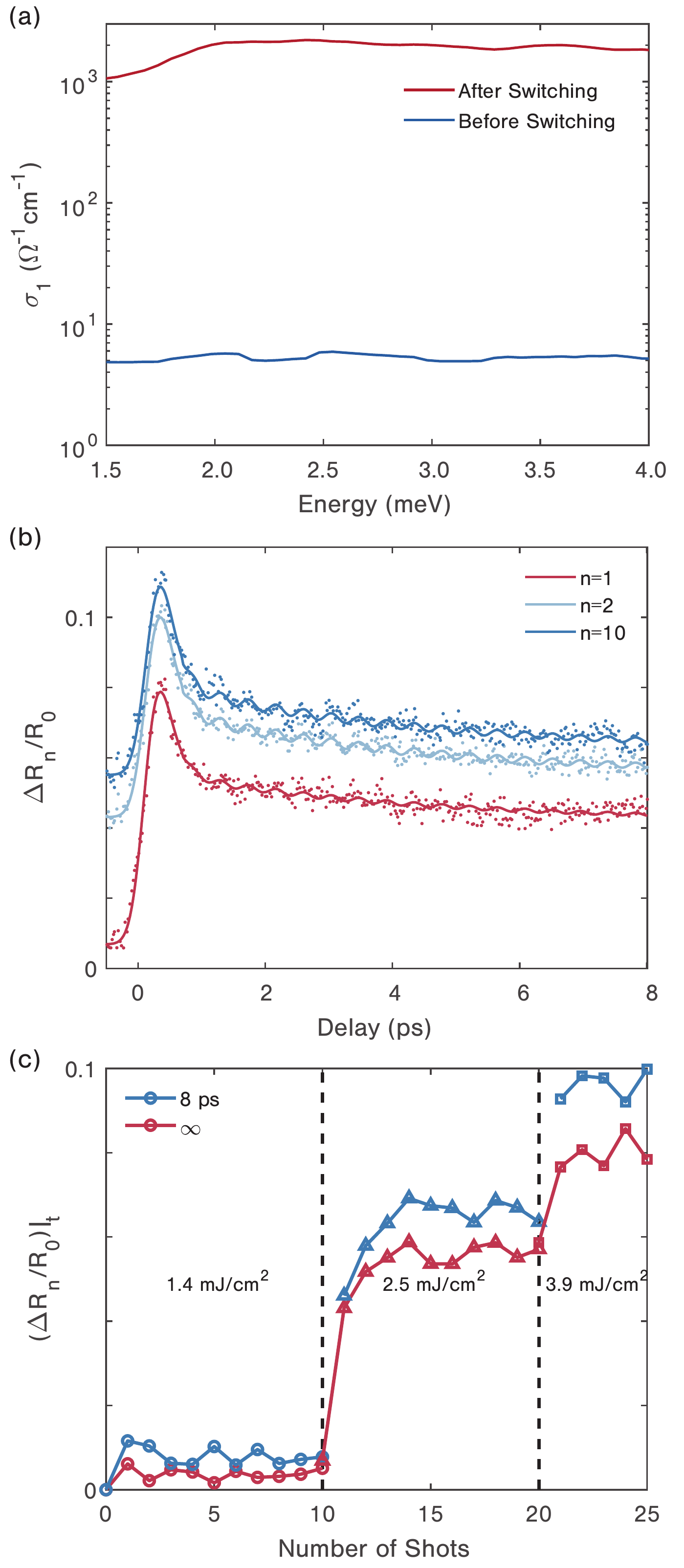}
		\caption{\label{fig:Fig2} Dynamics of the emergent hidden phase formed after photoexcitation in 1\textit{T}-TaS$_2$. (a)  Steady-state real part of the THz conductivity increases by two orders of magnitude following switching.  (b)  Shot-by-shot transient reflectivity measurements {(dots)} at 7.8 K following irradiation with 2.5 mJ/cm$^2$ 1.55 eV pump pulses are shown along with the corresponding fits to biexponential decays {and oscillatory responses} (solid lines). Ten shots were recorded at this fluence but only the first, second and tenth are reproduced here for clarity. Each irradiation was separated by more than	 10~s. (c)  The corresponding shot-to-shot change in steady-state reflectivity (red) and 8 ps reflectivity (blue) over the full sequence of irradiations. Note that $n = 1$ in (b) corresponds to the 11th shot in  (c). }
	\end{figure}
	
	With the formation of the H state established, we seek to address key questions about its underlying dynamics. The first unknown regards the timescale for H state nucleation. To measure this, we use single-shot transient reflectivity, with both pump and probe pulses centered at 1.55 eV. Figure 2(b) shows the data acquired at 7.8 K upon sample irradiation with one, two, or ten consecutive pump pulses (keeping for each one the same incident fluence used in the THz experiment of Fig. 2(a)). Upon absorption of the first pump pulse, we observe a {time-resolution-limited} increase in the sample reflectivity, followed by a sub-picosecond relaxation and a slower decay over a few picoseconds (red dots). The {relaxation} dynamics are well captured by a biexponential fitting function, with relaxation {times} of 0.17 ps and 3.7 ps, as shown by the solid red curve. For long time delays ($> 8$ ps), the slow decay converges to a large ($\sim5$\%) and stable reflectivity offset that never recovers back to the original, pre-time-zero value. This new value of the NIR reflectivity provides a distinct signature of H state formation, this time at photon energies as large as a few eV. We also observe that the process of H state writing/erasing is completely reversible: cycling the sample temperature between 7.8 K and 80 K erases the H state and restores the pristine C state (see Supplementary Fig. S2). Upon cooling down to 7.8 K, the H state can be created again by applying another laser pulse at 2.5 mJ/cm$^2$ fluence, resulting in a transient reflectivity curve identical to the red trace in Fig. 2(b). { After the first shot, an additional shot at the same 2.5 mJ/cm$^2$ fluence induced a further small change in the static reflectivity, as shown in Fig. 2(c) (shots 11 and 12; the first ten shots are at 1.4 mJ/cm$^2$ fluence). Subsequent shots at 2.5 mJ/cm$^2$ fluence (shots 13-20) resulted in no further systematic change. See Supplementary Fig. S3 for the entire sequence. The increased static reflectivity value after the first shot differs only slightly from its value at 8 ps time delay after photoexcitation, and in fact there is only a modest change after 1 ps delay.  From this, we infer} that the stabilization of the H state is complete within several picoseconds of the first pump laser shot. 
	
	Having revealed the ultrafast dynamics of H state formation triggered by a single laser pulse, we obtain a broader overview of the switching behavior and evolution of the H state {through single-shot measurements at several} excitation {fluence} regimes. {Figure 2(c) shows} a sequence of 25 single-shot irradiations at three different fluences and for each shot compares the reflectivity change at 8 ps with the steady-state reflectivity of the sample long after the shot. We observe that irradiating the sample with one pump pulse at 1.4 mJ/cm$^2$ produces only a slight increase (max. 1\%) in the sample reflectivity at 8 ps and that this reflectivity signal decays over time. Further irradiation with a train of 9 pulses at the same fluence does not {result in significant} changes in the signal behavior. {As discussed above,} when the pump fluence exceeds a threshold value of $F_{th,7.8\text{K}}$ = 2.5 mJ/cm$^2$, the sample undergoes a sudden increase in steady-state reflectivity. Another sudden change occurs when the fluence is increased to 3.9 mJ/cm$^2$, an observation that is consistent with the photoinduced metallicity \cite{sun_hidden_2018}. {At this fluence we see once again that} the full extent of switching is induced by the first shot, and further shots at the same fluence have {no further systematic effects.}  {As in the 2.5 mJ/cm$^2$ fluence case, the steady-state reflectivity value reached after the first each shot differs by only a modest amount from the value at 8 ps (or 1 ps) delay following excitation,} which indicates that switching {at this fluence also} occurs on a picosecond timescale.  We note also that at all fluences, including the first shots at the two high fluences, the relaxation kinetics show a subpicosecond decay component followed by a slower multi-picosecond decay. (The values are indicated in Supplementary Table S1) The decay rates at 1.4 mJ/cm$^2$ and 2.5 mJ/cm$^2$ fluences are similar, even though no switching occurs at the lower fluence. At the highest fluence, the fast decay component is faster and the slow component is slower but the same trends seen at lower fluence hold (see Fig. S3(c)). From these results we attribute the first process to electronic relaxation due to typical electron-phonon interactions while the following processes are dominated by thermal fluctuations which involves the formation of disordered structures (see Section \ref{sec:theory}).  The consequences of these charge-order fluctuations are entirely different from the static thermal response which should result in a NIR reflectivity change of opposite sign to what is observed, i.e. a decrease in $\Delta R/R$.
	
	We also {gain insights into} the pathway through which the system enters the H state \cite{stojchevska_ultrafast_2014, ravnik_real-time_2018} by observing the CDW order parameter. The single-shot traces in  Fig. 2(b) and Fig. S3 show 2.4 THz coherent oscillations in the time domain, which are characteristic signatures of the CDW amplitude mode \cite{demsar2002femtosecond, toda_anomalous_2004, perfetti2008femtosecond, mann2016probing}. The amplitude of this collective mode becomes weaker as more of the sample switches into the H state, (first shots at 2.5 and 3.9 mJ/cm$^2$ fluences) indicating that the melting of the original CDW order is tied to the formation of the metastable state (see Supplementary Note 2 and Fig. S4). 
	
	\subsection{Averaged single-shot transient reflectivity of the transient H state}
	
	\begin{figure*}
		\includegraphics[width=0.9\textwidth]{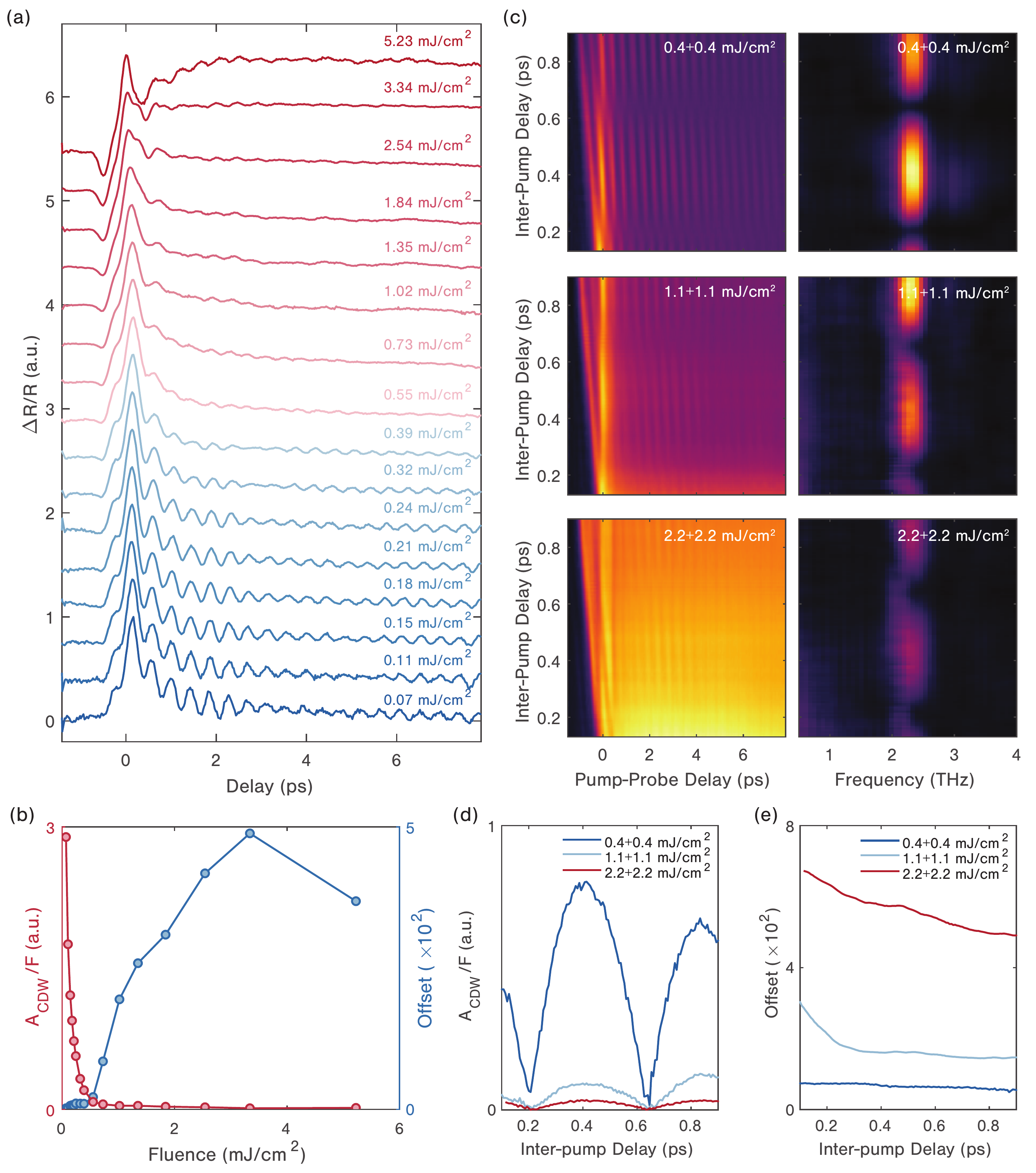}
		\caption{\label{fig:Fig3} Far-from-equilibrium NIR transient reflectivity of 1\textit{T}-TaS$_2$ at 80 K.  (a) Single-shot transient reflectivity measurements averaged over 100 shots at 80 K following photoexcitation with an 1.55 eV pump pulse over a wide range of excitation fluences. (b) The CDW amplitude mode strength normalized by the fluence (red) and the offset response (blue) extracted from the fits to (a) as a function of excitation fluence. At low fluence, the offset response is nearly zero and only appears above a critical threshold fluence of 0.55 mJ/cm$^2$, above which it rapidly increases with increasing fluence. For the CDW amplitude mode, the coherent response is strongest at low fluence and then drops steadily as the fluence is increased, nearly disappearing at the same threshold fluence. (c) Single-shot transient reflectivity measurements as a function of both inter-pump delays and pump-probe delays with three distinct fluence regimes: 0.4+0.4 mJ/cm$^2$ or low fluence regime, 1.1+1.1 mJ/cm$^2$ or medium fluence regime, and  2.2+2.2 mJ/cm$^2$ or high fluence regime.  (d) the amplitude mode strength and (e) offset extracted from the fits to (c) for the three fluence regimes as a function of inter-pump delay.  }
	\end{figure*}
	
	To resolve the dynamical behavior of the photoinduced phase transformation absent any persistent response, we raise the sample temperature to 80 K and conduct single-shot NIR transient reflectivity measurements. In this regime, we demonstrate that the lifetime of the H state is shortened to less than 20 ms. {The recovery is too slow to allow conventional pump-probe measurements at our 1-kHz laser repetition rate, but we can} track the CDW dynamics with high signal-to-noise ratio by averaging over many single-shot traces. We find that the threshold fluence for H state formation is reduced to $F_{th,80\text{K}} \sim$ 0.55 mJ/cm$^2$, i.e., a value that is significantly lower than $F_{th,7.8\text{K}}$. This suggests that the ground state is more susceptible to disruption from photoexcitation at the higher temperature compared to when the system is deeply trapped in the C state at the lower temperature.
	
	Figure 3(a) shows the measured single-shot spectroscopy data at various incident fluences, with each curve obtained from averaging 100 single-shot traces. Unlike in our conventional pump-probe measurements – in which sample damage occurs above 1.5 mJ/cm$^2$ (see Supplementary Note 3) – single-shot experiments enable us to enter a hitherto unexplored excitation regime (up to 5.23 mJ/cm$^2$) and provide a comprehensive view of the H state formation. Below the threshold fluence for the creation of the H state ($F < F_{th,80\text{K}}$), we observe that the sinusoidal signal due to the amplitude mode appears promptly after the initial electronic response and oscillates with a frequency of 2.4 THz. Increasing the incident fluence results in the weakening of the amplitude mode, indicating a transient suppression of the CDW order parameter.\cite{demsar2002femtosecond,baldini2020discovery} Figure 3(b) shows the strength of the amplitude mode, {$A_{CDW}$, and the $t > 8$ ps offset signal component} as a function of fluence (fitting procedures are described in Supplementary Note 5 and fits are shown in Figure S8). Once the fluence exceeds $F_{th,80\text{K}}$, the mode intensity reduces almost to zero, signaling a nearly complete melting of the CDW order. Simultaneously, an offset emerges, a response that resembles the dynamics involved in entering the persistent H state at 7.8 K. This unique feature is not observed in the conventional pump-probe data in either this or previously reported works \cite{ravnik_real-time_2018,ravnik2021time}, and is not compatible with a mere increase in the lattice temperature (see Supplementary Fig. S5 and Supplementary Note 4). Therefore, we can conclude that {the} offset is the fingerprint of the metastable H state at 80 K. Above $F_{th,80\text{K}}$, the offset increases monotonically with the excitation fluence, up to ~3.3 mJ/cm$^2$, before being suppressed at the highest fluence. Here, the time-domain trace also changes dramatically, with the signal upturning after the initial impulsive response and with a new oscillation emerging at a frequency of 2.1 THz. We ascribe the latter to the $E_g$ phonon mode.\cite{sugai_comparison_1981, toda_anomalous_2004} Unlike the amplitude mode, the $E_g$ phonon mode is not directly coupled to the instantaneous perturbation of the charge density, but rather to the periodic lattice distortion.\cite{toda_anomalous_2004} In this framework, the disappearance of the CDW amplitude mode and the persistence of the $E_g$ phonon {mode} at high fluence {suggest} that the melting of the CDW does not fully suppress the periodic lattice distortion \cite{eichberger2010snapshots,ikeda2019photoinduced}(See Supplementary Note 5 and Fig. S7-S8). 
	
	From this data set, we can also establish that the CDW melting leading to H state creation proceeds non-thermally. Indeed, at $F_{th,80\text{K}}$, the lattice temperature remains well below the equilibrium CDW melting temperature (see Supplementary Note 4).\cite{zhang_photoexcitation_2019} We further verify this by performing a series of double-pump-pulse  single-shot reflectivity experiments, varying the inter-pump delay while keeping other parameters unchanged (Fig. 3(c), left column). For non-thermal melting, one expects an increase in {the} offset and a further suppression of the amplitude mode oscillation as the inter-pump delay decreases, {since the system is effectively pumped by a single stronger pulse before it fully relaxes.} However, for thermal melting, no such variation would occur at short inter-pump delays (within several picoseconds) {since the total energy deposited by the pump pulse pair does not change significantly.} We explore three different excitation regimes for the pump pulses: low ($F < F_{th,80\text{K}}$), medium ($F > F_{th,80\text{K}}$), and high ($F \gg F_{th,80\text{K}}$) fluence. We then perform a Fourier transform analysis of the oscillatory signal along the pump-probe delay axis (Fig. 3(c), right column). At all fluences, we observe that the amplitude mode is modulated as the inter-pump delay is changed, manifesting itself as interference effects in the Fourier transform signal. However, as the fluence increases, the amplitude mode oscillations become less prominent. The offset response exhibits a strong dependence on inter-pump delay. The strengths of the amplitude mode and the offset for these three regimes are extracted from the Fourier transforms and the fits as functions of inter-pump delay (Fig. 3(d,e)). In the regime with medium fluence, the offset drops substantially when the inter-pump delay is increased, accompanied by the recovery of the amplitude mode strength after the second to third periods of delay. In the high fluence regime, the amplitude mode is further suppressed, and the long-lived offset becomes prominent, reflecting that the material enters the photoinduced metastable H state. These observations reaffirm that the melting is non-thermal, showing a strong dependence on the peak carrier density, and that a threshold density of initially absorbed photons is required to trigger the long-lived metastable state. 
	
	\subsection{Single-shot NIR-pump THz-probe signals}
	
	\begin{figure*}
		\includegraphics[width=0.9\textwidth]{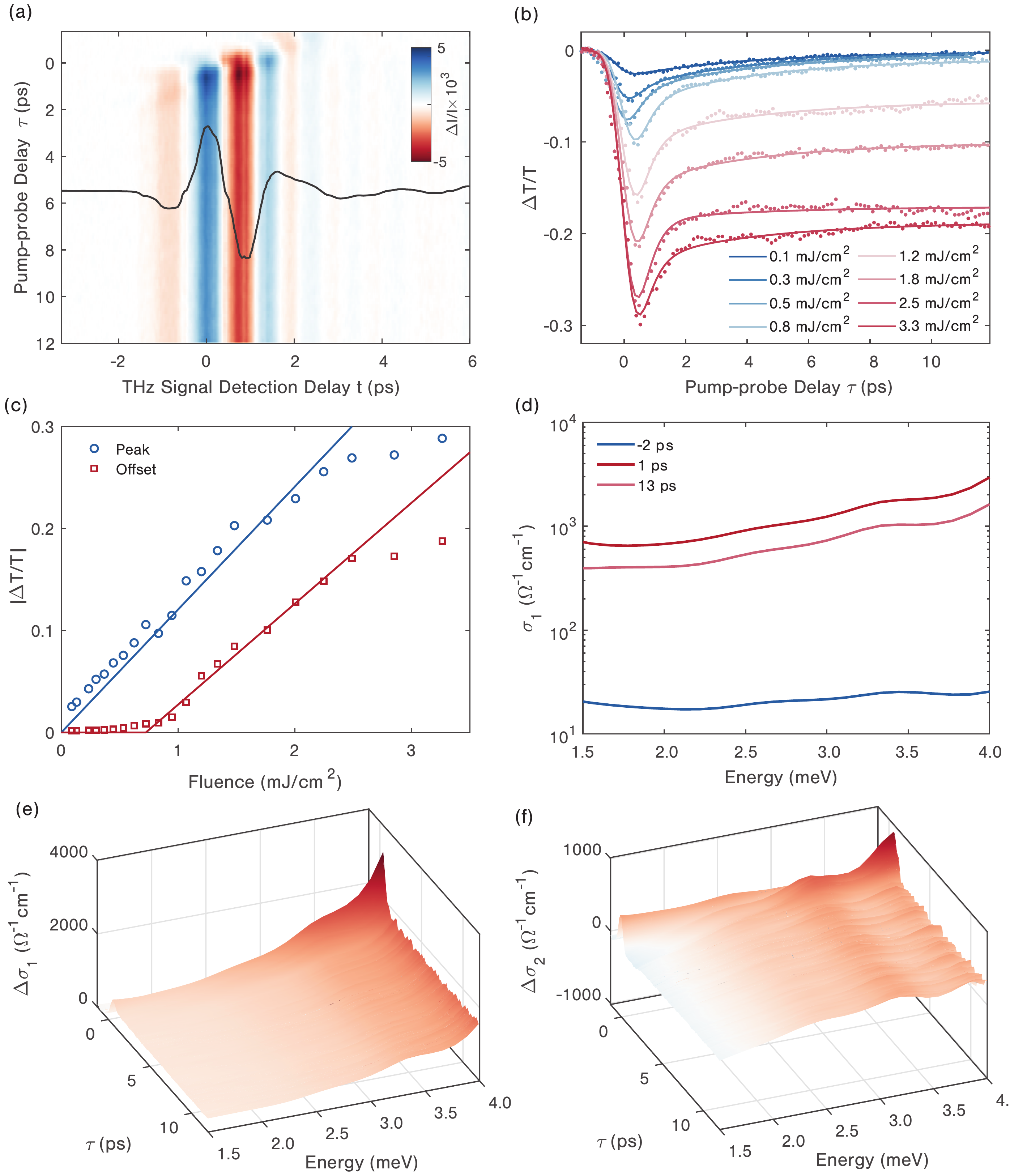}
		
		\caption{\label{fig:Fig4}
			Far-from-equilibrium THz conductivity of 1\textit{T}-TaS$_2$ at 80 K. (a) Photoinduced change in the THz electric field trace, $E(t)$, at 80 K as a function of pump-probe delay, $\tau$, with 1.55 eV nm pump and an excitation fluence of 3.3 mJ/cm$^2$. {The transmitted THz field profile without photoexcitation (black line) is sketched with the same horizontal time axis. In the data shown, the pump pulse decreases the transmission of THz light through the sample, with the effect persisting for the entire range of delay times following the pump pulse.} (b) Spectrally-integrated differential transmission of the main THz peak through the sample following photoexcitation over a large range of excitation fluences. (c) Peak response (blue) and offset response (red) extracted from the fits to (b) as a function of excitation fluences. The offset response only shows up above a critical excitation fluence and increases with fluence above the threshold. The solid lines {show roughly linear trends in the peak response and the above-threshold offset with respect to the excitation fluence up to about 2 mJ/cm$^2$ fluence.} (d) Comparison between the real part of the transient THz conductivity before and after photoexcitation of 3.3 mJ/cm$^2$. There is an increase of two orders of magnitude in the transient conductivity following photoexcitation. (e,f) The spectro-temporal evolution following the photoinduced change with fluence of 3.3 mJ/cm$^2$ (same as (a)) in the real and imaginary parts of the THz conductivity. }
	\end{figure*}

	Finally, we establish how the initial CDW state transforms into the H state by investigating the pump-induced changes to the THz conductivity with single-shot THz spectroscopy, a recently developed technique \cite{teo2015invited, teitelbaum_dynamics_2019} for measuring long-lived low-energy responses (details are given in the Methods section). The THz trace is read out by 400 electro-optic sampling probe pulses that arrive at the sample at different delay times relative to the pump laser pulse, allowing us to collect the full time-dependent conductivity response on each shot. In the present study, we repeated the measurements at a repetition rate of 50 Hz, which still allows a high total data acquisition rate given the number of time points collected on each shot. This also ensures that the observed dynamics {arise} from the effect of each individual pump pulse rather than any cumulative effects of multiple pulses {(no variation in signal was seen at still lower repetition rates)}. Figure 4(a) shows the photoinduced change in the transmitted THz field, $E(t)$, at 80 K as a function of pump-probe delay, $\tau$, under the excitation fluence of 3.3 mJ/cm$^2$. By analyzing the change in the spectrally-integrated THz field transmission ($\int |E(\omega)| d\omega$), we can extract the reduced THz transmission after photoexcitation, presented in Fig. 4(b) for a large range of excitation fluences. Also shown in solid lines are fits to biexponential decays convoluted with the instrument response function. Although the time resolution is limited by the THz pulse duration, we observe a response that is reminiscent of the NIR transient reflectivity traces,  exhibiting similar timescales and changing significantly above a critical excitation threshold around $F_{th,80\text{K}}$. At larger fluences, a long-lived offset dominates after the initial resolution-limited response  and the following relaxation process. We then plot the offset and peak transmissivity retrieved from the fits in Fig. 4(b). As shown in Fig. 4(c), the peak amplitude displays an approximately linear scaling as a function of excitation fluence (especially in the low fluence regime), while the offset signal shows an increasing trend only above $F_{th,80\text{K}}$. Figures 4(e) and 4(f) display the spectro-temporal evolution of the real ($\Delta\sigma_1$) and imaginary ($\Delta\sigma_2$) parts of the differential THz conductivity following excitation with a single 3.3 mJ/cm\textsuperscript{2} pump pulse. Upon laser excitation, $\Delta\sigma_1$ is increased by two orders of magnitude (Fig. 4(e)) and $\Delta\sigma_2$ becomes negative at low energies. For longer pump-probe delay, both  $\Delta\sigma_1$ and $\Delta\sigma_2$ are compatible with a Drude-Smith type of response, similar to the steady-state conductivity when the H state is switched on at 7.8 K.  This response is typical of confined carriers affected by strong backscattering.  In this case, scattering likely occurs at either the boundaries of the metallic domains of the H state or the boundary between the photo-switched H state and the pristine C state, which results from the limited penetration depth of the 1.55 eV pump beam.

	\section{Theory of pump-induced dynamics} \label{sec:theory}
	\begin{figure}
	\includegraphics[width=\linewidth]{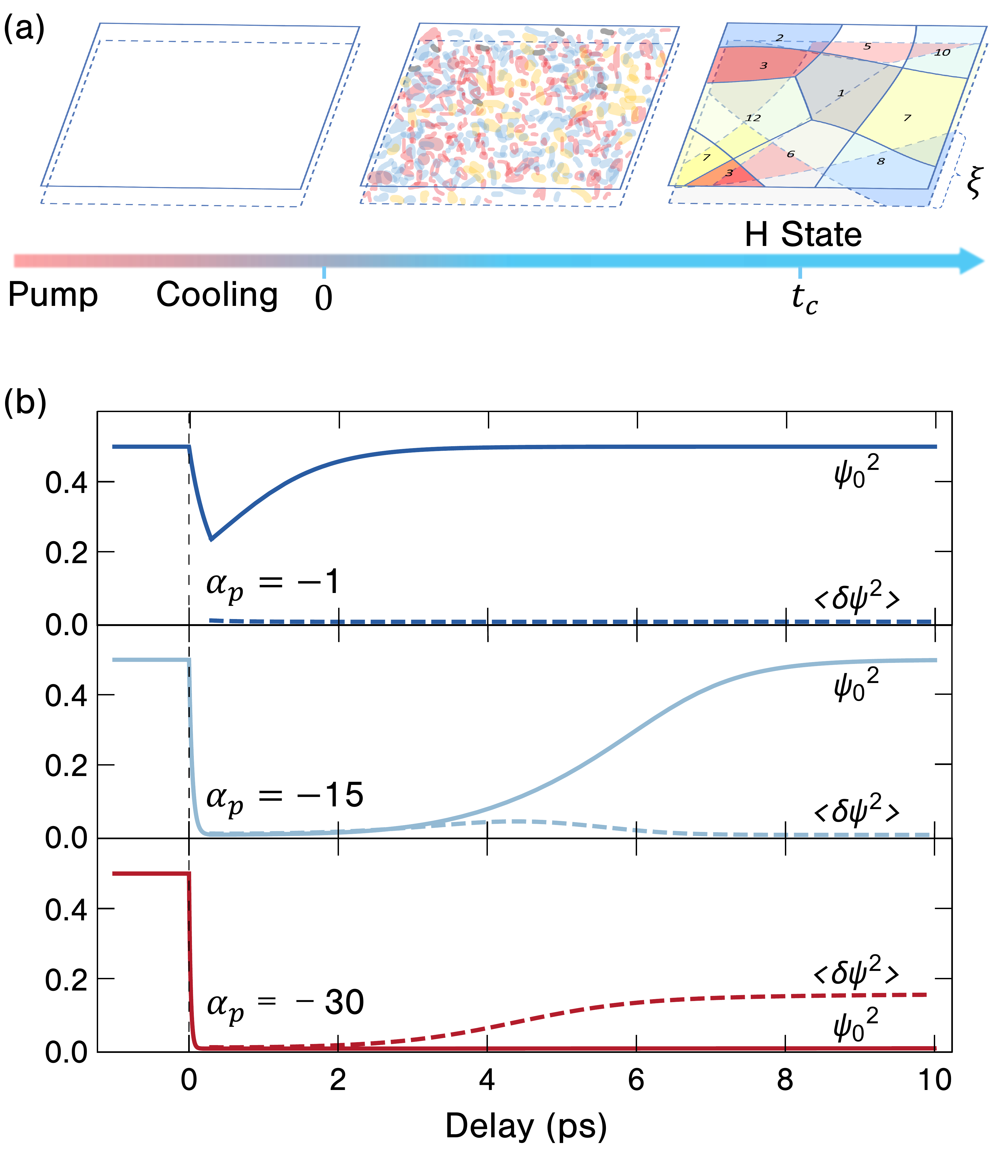}
	\caption{
	(a) Schematic illustration of the pump induced dynamics of the sample. The horizontal arrow is the arrow of time following the pump pulse. The two planes represent two adjacent 1$T$-TaS$_2$ layers. Each number labels a domain in the H state corresponding to one of the $13$ possible low energy CDW states.
	(b) The time evolution of the square of the mean field order parameter and its fluctuations computed from Model A (\equa{eqn:TDGL}). The effect of the pump is modeled as a time-dependent quadratic term coefficient $\alpha(t)=\alpha_0+(\alpha_p-\alpha_0)\left[\Theta(t)-\Theta(t-t_p)\right]$ where $\alpha_0=1$ is its static value. The pump time is chosen as $t_p=0.3 \unit{ps}$ since the pump duration is $0.1 \unit{ps}$ while the electronic system needs approximately $0.2 \unit{ps}$ to cool down. The Ginzburg number and the intrinsic CDW relaxation time is the same as \equa{eqn:ginzburg}. Three different values of $\alpha_p$ ($-1$, $-15$ and $-30$) are chosen to represent three different pump fluence regimes (low, medium and high).
	}
	\label{fig:H_state}
	\end{figure}
	The observation of the metastable state and the melting of the CDW order suggests that the formation of the H state can be described by the mechanism of fluctuation explosion~\cite{sun.2020.competing_order} which follows the non-thermal collapse of charge order. We note that thermal fluctuations have  been found to play important roles in the pump-induced dynamics of other CDW systems as well~\cite{Zong.2019,Zong:2019tx,Kogar:2020vt, Dolgirev.2020,Zong.2021}.
	
	The pump induced dynamics of the CDW order can be described by the space-time dependent order parameter field $\psi(\mathbf{r},t)=\psi_0(t)+\delta \psi(\mathbf{r},t)$ on each two dimensional layer of 1$T$-\ce{TaS2}. The free energy is in the usual Landau form:
	\begin{align}
		F/E_c=-\alpha \psi^2 + \xi_0^2 (\nabla \psi)^2 + \psi^4
		\label{eqn:F}
	\end{align} 
	where $E_c$ is the condensation energy density, $\xi_0$ is the bare coherence length, and $\alpha(T)$ is $O(1)$ at zero temperature $T$. Since there are $13$ different stable free energy minima corresponding to aligning an arbitrary lattice site with the  $13$ different Ta atoms in the David star, the order parameter cannot be represented by a simple real number. However, the discrete symmetry means that its scaling behavior can be equivalently described by the Ising order parameter with two minima in \equa{eqn:F}.
	
	To describe the dynamics induced by a strong pump above the switching threshold, we assume that  the order parameter fields evolve according to relaxational (or ``Model A''~\cite{Hohenberg1977}) dynamics defined by a free energy functional $F$: 
	\begin{align}
		\frac{1}{\gamma}\partial_t \psi(\mathbf{r},t)
		=
		-\frac{1}{E_c}\frac{\delta F(t)}{\delta \psi (\mathbf{r},t)} + \eta(\mathbf{r},t)
		\,
		\label{eqn:TDGL}
	\end{align} 
	where $\gamma=1/\tau$ is the intrinsic relaxation rate. We make the natural assumption that the applied NIR pump field raises the temperature of the electronic system such that $\alpha(T(t))$ and thus $F(t)$ are time dependent. $\eta$ is the random noise satisfying the fluctuation-dissipation theorem
	$
	\left<\eta_i(\mathbf{r},t)\eta_i(\mathbf{r}^\prime, t^\prime\right>=\frac{2T(t)}{\gamma_i E_c} \delta(\mathbf{r}-\mathbf{r^\prime},t-t^\prime) 
	\label{etacorrelator}
	$.
	
	In this framework, our results can be explained in the picture shown in \fig{fig:H_state}. A strong pump pulse transiently heats up the electronic system and renders $\alpha$ negative such that the mean field CDW order parameter $\psi_0$ decays to a negligible value, even smaller than its thermal fluctuations. After the pump, the electron system quickly cools down by transferring its excess energy to the lattice within a time scale of $0.3 \unit{ps}$. Then, the CDW fluctuations $\delta \psi(\mathbf{r},t)$ grow exponentially, forming random domains~\cite{sun.2020.competing_order} within each layer. Since the interlayer coupling is much weaker than the intralayer one, we treat each layer individually such that the random domains of adjacent layers are not correlated. Giving any two overlapping domains on adjacent layers, there is very little chance that they are perfectly aligned to form the original insulating C state (center of the  David star on top of each other \cite{ritschel2015orbital,stahl2020collapse}). Therefore, one has an H state of random domains, which is mostly metallic due to this altered stacking order. \cite{ritschel2015orbital, stahl2020collapse, Ma:2016wn} As shown in \fig{fig:H_state}(b), weak pump pulses can transiently quench the CDW order, $\psi_0^2$, either partially (e.g. $\alpha_p=-1$) or almost completely (e.g. $\alpha_p=-15$), but do not result in a lasting fluctuation of the order parameter, $\langle\delta\psi^2\rangle\sim0$, consistent with the low fluence regime of our experiments. At high fluences (e.g. $\alpha_p=-30$), the thermal fluctuations grow exponentially, which results in a long-lived disordered metastable state that forms within roughly 6 ps.  Successive irradiations at the same fluence would induce the same magnitude of thermal fluctuations and thereby contribute no further changes to the degree of H state formation, in agreement with our experimental observations in Fig. 2(b). 
	
	In principle, there are driving forces to relax this H state back to the C state: the weak interlayer coupling provides a force to shift the local CDW horizontally to align the overlapping domains, and the intralayer dynamics of \equa{eqn:TDGL} tend to shorten the total domain wall length to reduce the energy. These relaxation mechanisms account for the transience of the H state discovered at high temperature; however, there are several reasons why this metastable H state is long-lived or even persistent at low-temperature. Firstly, due to frustration of domain overlap on adjacent layers, the intralayer restoring force tends to cancel out. Secondly, the domain walls may be pinned by disorder, which provides barriers for their motion and enhances the lifetime of the H state exponentially as the temperature is lowered.
	
	From a theoretical prospective, our single-shot measurements also provide a benchmark that enables us to extract several fundamental parameters of the Landau-Ginzburg-Wilson theory from the system. In the fast-cooling limit (electron-phonon scattering time shorter than the order parameter growth time $t_c$), the time needed to form the metastable H state and the domain size are \cite{sun.2020.competing_order} 
	\begin{align}
		t_c \sim \frac{\tau}{4\alpha} \ln \frac{1}{\zeta}
		,\quad
		\xi \sim \xi_0 \sqrt{2/\alpha}  \sqrt{\ln \frac{1}{\zeta}}
		\,
	\end{align}
	where $\zeta$ is the Ginzburg parameter (or `Ginzburg number') that determines the accuracy of the mean field picture. From the STM images of the H states induced by optical and electrical pulses~\cite{ravnik2021time, Ma:2016wn, Park:2021va}, the bare coherence length (roughly the width of the domain walls at a temperature much lower than $T_c$) is about $\xi_0=3 \unit{nm}$ while the domain size is about $\xi=10 \unit{nm}$. Together with the new observation that the evolution time is $t_c=3.7 \unit{ps}$, we obtain the Ginzburg parameter and the intrinsic relaxation time of the in-plane CDW:
	\begin{align}
		\zeta \sim 4 \times 10^{-3}
		,\quad
		\tau \sim 2.7 \unit{ps}
		\,.
		\label{eqn:ginzburg}
	\end{align}
	The small value of $\zeta$ means that the CDW is a weakly correlated state, meaning the theory used here is reasonable. \color{black}
	
	\section{Discussion}
	
	Taken together, our single-shot time-resolved measurements combined with the stochastic time-dependent Ginzburg-Landau theory establish a fundamental connection between the transient and persistent H states in 1\textit{T}-TaS$_2$. Their similarities in dynamics after photoexcitation show that H state formation follows the same pathway, regardless of the transient or persistent nature of the resulting phase. In both regimes, we observe direct evidence for the ultrafast nature of photoinduced metallicity. Upon above-threshold laser excitation, the CDW order melts on a non-thermal timescale faster than our experimental resolution. After the collapse of the electronic order, instead of returning to the original C state, thermal fluctuations~\cite{sun.2020.competing_order} dominate the following relaxation process, allowing the metastable H state to set in on a timescale of several picoseconds. Such a fast timescale characterizing the stabilization of the H state is in stark contrast to the timescale involved in other photoinduced persistent insulator-to-metal transitions, for example when the formation of the conductive states relies on slowly developing ferromagnetic domains \cite{teitelbaum_dynamics_2019}. Our findings support a picture {in which} the ultrafast collapse of a charge gap can lead to a lasting change in electronic structure and modification of the free-energy landscape, as confirmed by theory.  In this case, the primary difference between the two temperature regimes is a kinetic one, with the eventual relaxation of the H state and the reformation of the C state being faster at high temperature. It should be noted that our findings prompt comparison with previous results that have inferred the timescale of H state formation at much higher temperatures (e.g. $> 80$ K) with a three-pulse stroboscopic technique \cite{ravnik_real-time_2018, ravnik2021time}. In those measurements,  the correspondence of this transient state with the persistent one was unclear and no comparison of their photoinduced dynamics was available. Direct tracking of single-shot switching events without introducing additional pump pulses  clarifies the formation of the H state under conditions directly comparable with previous reports of the photoinduced metallicity in 1\textit{T}-\ce{TaS2}\cite{stojchevska_ultrafast_2014,sun_hidden_2018} and yields crucial information not accessible otherwise.  Along these lines, our observations show distinct optical responses for the formation of the H state (i.e., the concomitant {increases} in NIR reflectivity and THz conductivity) which demonstrate that the H state is substantially different from the NC state, and in general from any other equilibrium states. These results highlight the potential to utilize the functionality  of hidden states on-demand and over a wide range of timescales, providing additional control for future optoelectronic devices based upon ultrafast photoresponses. More fundamentally, our single-shot techniques can be used to study other materials with photoinduced metastable phase transitions that remain unexplained, such as systems with multipolar orders \cite{xu2020spontaneous}, ferroelectric superlattices \cite{stoica_optical_2019}, and molecular solids with light-induced superconductivity \cite{mitrano_possible_2016, budden_evidence_nodate}.
	
	\appendix
	\section*{Methods}
	
	\subsection{Single crystal synthesis}
	Single crystals of 1\textit{T}-\ce{TaS2}  were synthesized via a chemical vapor transport process \cite{inada1979hall}. The process we adopted here has been described elsewhere \cite{zong2018ultrafast}. We first mixed Ta powder (Alfa Aesar, 99.97\%) and S pieces (Sigma Aldrich, 99.998\%) with an off-stoichiometric ratio of 1:2.02; the mixture was placed in an evacuated quartz tube and kept at 970$^\circ$C for 2 days, and subsequently quenched in cold water. A second evacuated quartz tube containing the resulting \ce{TaS2} along with added \ce{I2} was placed in a temperature gradient from 920$^\circ$C (source) to 820$^\circ$C (sink) for two weeks. At the end of the growth, the quartz tube was quenched in water. Powder X-ray diffraction was performed to confirm that the resulting crystals are of the 1\textit{T}-\ce{TaS2} phase.
	
	\subsection{Single-shot NIR transient reflectivity}
	Single-shot dual echelon NIR transient reflectivity measurements were conducted by using the output of a Ti:Sapphire laser system (800 nm, 1 kHz, 3 mJ, 70 fs pulses) and down-counting to 10 Hz via a Pockels cell and polarizing beamsplitter. A mechanical shutter synchronized to the laser output was used to select individual laser pulses. The majority of the laser power was sent into a modified Michelson interferometer which generated two delayable pump pulses. Approximately 100 $\mu$J of the laser output was split off and passed through dual echelons, resulting in a temporally offset $20\times20$ grid of pulselets spanning about 9.3 ps. These pulselets were then focused onto the sample along with the pump pulse (spot size $\sim$ 600 $\mu$m) in a transient reflectivity geometry that included a reference grid of beamlets (not shown in Fig. 1) for normalization of the signals as described previously \cite{shin2014dual}. A \textit{4f} imaging systems was used to relay the signal and reference grids onto a  CCD (Hamamatsu Orca-ER). These grids were then combined, binned and unfolded to generate a pump-probe transient reflectivity trace measured in each laser shot \cite{teitelbaum_real-time_2018,shin2014dual}.   \\
	
	\subsection{Single-shot optical pump THz probe spectroscopy}
	The laser output was down-counted to 50 Hz {and no sample damage was observed in the measurement}. 1 mJ of the laser output was used to generate THz probe pulses via optical rectification in \ce{LiNbO3} using a tilted pulse front geometry. The remaining $\sim$ 1 mJ was divided between optical pump and electro-optic sampling beams. The THz probe and optical pump were focused onto the sample with spot sizes of roughly 1 mm and 2 mm respectively. Single-shot readout of the THz waveform was performed by passing the electro-optic sampling beam through dual echelons and then focusing the resulting grid of beamlets and THz onto a 1 mm GaP crystal. These grids were then imaged onto a CCD (Andor Zyla) after being split into perpendicularly polarized images by passing through a quarter waveplate and a Wollaston prism as shown in Fig. 1. The final THz waveform was obtained by binning, unfolding and then subtracting the two resulting time traces. To measure the two-dimensional optical pump-THz probe maps, the THz probe and optical pump were chopped at 25 Hz and 12.5 Hz respectively. The single-shot THz readout method has been described in detail previously \cite{teo2015invited, teitelbaum_dynamics_2019}. 
	
	\noindent The THz conductivity of the photoexcited region was obtained from the experimentally determined optically induced THz transmission spectra, $T(\omega)$, using the Tinkham equation \cite{glover1957conductivity}:
	\begin{equation*}
		\tilde{\sigma}(\omega) = \frac{n_s+1}{Z_0d}\left(\frac{1}{\tilde{T}(\omega)+1}\right),
	\end{equation*}
	where $n_s$ is the index of the substrate, $Z_0$ is the impedance of free space and \textit{d} is the sample thickness or the optical penetration depth. In the data analysis, we accounted for the penetration depth mismatch between the NIR pump and THz frequencies by treating the optically excited region as its own sample with thickness commensurate with the optical penetration depth, i.e., $d = 45$ nm. {The overall thickness of the sample used in the THz transmission measurements was $\sim$ 50 $\mu$m.}

	\begin{acknowledgments}
	We thank Junbo Zhu for helping with the samples and for useful discussions. F.Y.G., Z.Z., Y.-H.C., Z.L. and K.A.N acknowledge support from the U.S. Department of Energy, Office of Basic Energy Sciences, under Award No. DE-SC0019126. L.Y. and J.G.C. acknowledge support from the Gordan and Betty Moore Foundation EPiQS Initiative (Grant GBMF3848 and GBMF9070).
	\end{acknowledgments}
	
	\bibliography{paper}
	
\end{document}